\documentclass[iop]{emulateapj}
\usepackage{times}
\usepackage{amssymb}
\usepackage{times}
\usepackage{txfonts}
\usepackage{bm}
\usepackage{color}
\usepackage{float}

\shorttitle{Line driven disk winds in Active Galactic Nuclei: 
The critical  importance of multiple scattering}
\shortauthors{Higginbottom et al.}

\begin{document}

\title{Line-driven Disk Winds in Active Galactic Nuclei: The Critical Importance of Ionization and Radiative Transfer}

\author{Nick~Higginbottom}
\affil{School of Physics and Astronomy, University of Southampton, Highfield, Southampton, SO17 1BJ, UK}
\email{nick\_higginbottom@fastmail.fm}
\author{Daniel~Proga}
\affil{Department of Physics \& Astronomy, University of Nevada, Las Vegas, 4505 S. Maryland Pkwy, Las Vegas,
NV 89154-4002, USA}
\author{Christian~Knigge}
\affil{School of Physics and Astronomy, University of Southampton, Highfield, Southampton, SO17 1BJ, UK}
\author{Knox~S.~Long}
\affil{Space Telescope Science Institute, 3700 San Martin Drive, Baltimore, MD 21218, USA}
\author{James~H.~Matthews}
\affil{School of Physics and Astronomy, University of Southampton, Highfield, Southampton, SO17 1BJ, UK}
\and
\author{Stuart~A.~Sim}
\affil{School of Mathematics and Physics, Queens University Belfast, University Road, Belfast, BT7 1NN, Northern Ireland, UK}

%\date{\today}

%\pubyear{2013}

%\newcommand\plotone[1]{\centering\includegraphics[width=\hsize]{#1}}
%\newcommand\plottwo[2]{\centering\includegraphics[width=0.45\hsize]{#1}\hfil\includegraphics[width=0.45\hsize]{#2}}
\newcommand{\ltappeq}{\raisebox{-0.6ex}{$\,\stackrel
{\raisebox{-.2ex}{$\textstyle <$}}{\sim}\,$}}
\newcommand{\gtappeq}{\raisebox{-0.6ex}{$\,\stackrel
{\raisebox{-.2ex}{$\textstyle >$}}{\sim}\,$}}

%\begin{document}
%\maketitle

\label{firstpage}

\begin{abstract}
Accretion disk winds are thought to produce many of the characteristic
features seen in the spectra of active galactic nuclei (AGN) and
quasi-stellar objects (QSOs). These outflows also 
represent a natural form of feedback between the central supermassive
black hole and its host galaxy. The mechanism for driving
this mass loss remains unknown, although radiation pressure mediated
by spectral lines is a leading candidate. Here, we calculate the
ionization state of, and emergent spectra for, the hydrodynamic
simulation of a line-driven disk wind previously presented by \cite{pk04}.
To achieve this, we carry out a comprehensive Monte
Carlo simulation of the radiative transfer through, and energy exchange
within, the predicted outflow. We find that the wind is much
more ionized than originally estimated.  
This is in part because it is much more difficult to shield any wind regions effectively 
when the outflow itself is allowed to reprocess and redirect ionizing photons. 
As a result, the calculated spectrum that would be observed from this particular
 outflow solution would not contain the ultraviolet spectral lines that are 
 observed in many AGN/QSOs. Furthermore, the wind is so highly ionized that
 line-driving would not actually be efficient. This does not necessarily mean that
 line-driven winds are not viable. However, 
 our work does illustrate that in order  to arrive at a self-consistent model of line-driven 
 disk winds in  AGN/QSO, it will be critical to include a more detailed treatment 
 of radiative transfer and ionization in the next generation of hydrodynamic simulations.
 \end{abstract}

%\begin{keywords}
\keywords{accretion, accretion disks - galaxies: active - methods: numerical - quasars: general - radiative transfer}
%\end{keywords}

\section{Introduction}
\label{introduction}
\begin{figure*}
\includegraphics{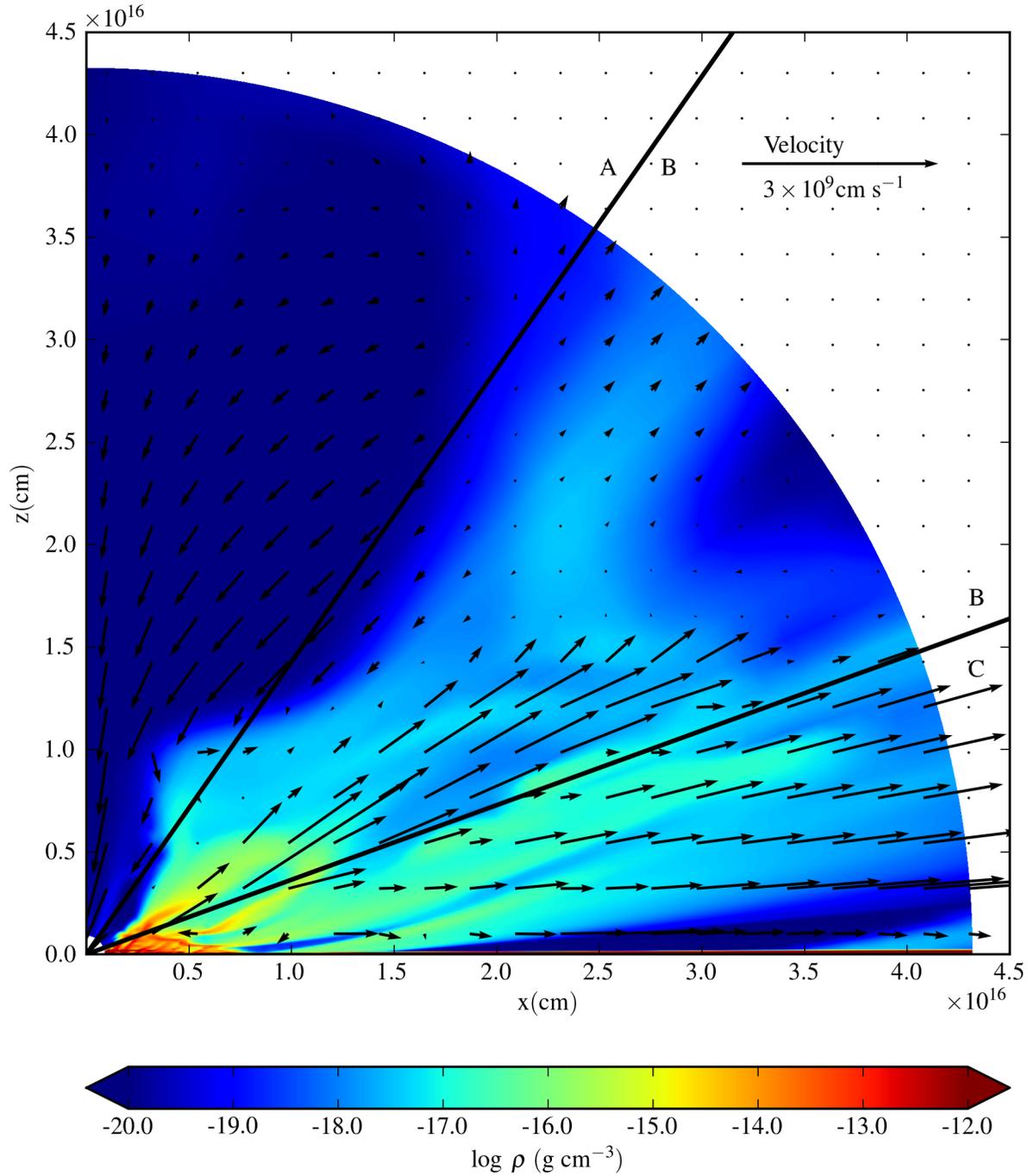}
\caption{The density (colours) and poloidal velocity (arrows) structure of the PK04 model. Radial lines delineate the
three zones described in the text. }
\label{lin_den}
\end{figure*}

Outflows are found in a vast range of accreting astrophysical objects,
from protostars to active galactic nuclei (AGN). In all of these
settings, quantifying the mass and energy flows involved is key to
understanding how the accreting objects evolve and interact with
their local environment. In the case of AGN, the observed outflows can
be split into two main classes: highly collimated, relativistic jets, and slower moving ($v \ltappeq
0.2c$), but more massive `disk winds' driven from the surface of the
accretion disk surrounding the central supermassive black hole. These
disk winds have been proposed as the underlying structure responsible
for many observed AGN spectral features, including the broad
absorption lines seen in a significant proportion of
QSOs \citep[e.g.][]{knigge08}, the so called broad absorption line
quasars (BALQSOs). However, the geometry of these winds, and even the 
mechanism by which they are launched, are still not known.

Several mechanisms have been proposed to produce the accelerating
force for disk winds in AGN, including gas/thermal pressure 
\citep[e.g.][]{weymann_82, begelman_91,
begelman_83,krolik_kriss}, magnetocentrifugal forces 
\citep[e.g.][]{blandford_payne_82, pelletier_pudritz} and radiation
pressure acting on spectral lines (``line-driving'')
\citep[e.g.][]{shlosman_85,murray_95}. All of these mechanisms have
been shown to produce outflows in suitable conditions. However, line driving
is particularly attractive, since the requisite force in this case
arises naturally when (primarily ultraviolet [UV]) photons produced by the
central engine are scattered by strong resonance lines. Many such 
lines are observed directly in the UV spectra of (BAL)QSOs, so line
driving must certainly be acting in some measure. Moreover, if line
driving is the dominant acceleration mechanism, it can produce a
unique signature in the profiles of absorption lines produced in the
wind (the so-called ``ghost of $\rm{Ly\alpha}$'', e.g.
\citealt{arav_95,arav_96}), and this may already have been seen
in several BALQSOs (\citealt{north_knigge_goad}, but also see
\citealt{cottis_10}).

Despite these circumstantial reasons for favouring line-driving as the
mechanism for the production of disk winds in AGN, there is one main
challenge to the model. For line-driving to be efficient, the
accelerated material has to be in a moderately low ionization state,
despite its proximity to the intense X-ray source at the center of the
accretion disk. If the gas becomes over-ionized, the ionic species
that can most effectively tap into the momentum of the radiation field
are simply not present. 

\cite{murray_95} proposed one solution, suggesting that 
`hitchhiking gas', material interior to the main outflow and itself
accelerated by pressure differences can shield the line-driven wind
from the central X-rays. 
Simulations such as those presented in \cite{proga_stone_kallman} 
and  \citet[][hereafter PK04]{pk04} also
produce a shield, but in this case via a `failed wind'. This failed
wind arises close to the X-ray-emitting region and quickly becomes
over-ionized as it rises above the disk surface. As it falls back, it
produces a dense shield that prevents the intense X-ray radiation from
reaching the outer parts of the outflow in its shadow. These 
low-ionization regions are then able to interact with UV radiation and
give rise to a strong line-driven wind. The same type of shielding
structure has also been seen in other models of disk winds
\citep[e.g][]{ris_elv,nomura_13}.

PK04 used hydrodynamical simulations to investigate whether a
line-driven outflow could be accelerated to high velocities purely 
by the radiation field produced by the accretion disk. They estimated 
the ionization state and temperature of the wind, taking into account
only the central source of ionizing radiation, attenuated by electron
scattering. Their main result was that the ionization state of the
wind behind the failed wind region remained low enough to permit line
driving, so that a fast, dense outflow was generated.  Absorption 
line profiles calculated from their wind model resembled
those observed
in BALQSOs (e.g. see Fig. 2 in in \citealt{prog_kuro_10}). It
is worth noting that PK04's wind, and especially its base, are
virialized systems. This is because a line-driven wind accelerates slowly, so
the rotational velocity is dominant over a relatively large distance while
the wind base is Keplerian and very dense. This means
that, if the broad line region (BLR) is associated with such a wind, 
black hole (BH) mass estimates based upon the assumption 
that the BLR is virialized remain correct \citep{kashi_13}.

The PK04 wind model has already been subjected to two more detailed
`post-processing' radiative transfer calculations. Both of these were
concerned with the impact of the predicted outflow on the X-ray
spectra of AGN/QSO. First,
\citet{schurch_09} performed a 1-D simulation which found that
the wind produced observable X-ray spectral features. Second,
\citet[][hereafter SP10]{sim_proga_10} carried out a multi-dimensional
radiative transfer simulation in which the ionization state and 
temperature structure of the wind were self-consistently
computed. They confirmed that the wind was able to imprint a variety
of characteristic features into the X-ray spectra of AGN. However, 
they also showed that scattering in the (failed) outflow is critically 
important in setting the ionization state of the wind in regions that
would otherwise be shielded from the central engine.

PK04 used a simplified treatment of radiative transfer and ionization
in their hydrodynamical simulations. These simplifications were
essential to make the simulations computationally feasible. However,
the work of SP10 implies that a careful treatment of radiative transfer
may be required in order to obtain a reliable estimate of the
ionization state of the wind. Since the ionization state, in turn,
determines the efficiency of the line-driving mechanism itself, it is
clearly important to check whether the wind model calculated by PK04
could actually maintain an ionization state that is consistent with
efficient line driving and with the production of broad UV absorption
lines. 

Here, we therefore carry out a full, multi-dimensional radiative
transfer calculation for the PK04 model that allows us to predict the
ionization state, temperature structure and emergent UV spectra for
this outflow. This extends the work done by SP10  to the longer 
wavelengths and lower ionization stages that are critical to allow 
effective line-driving and the formation of BALs. 
In particular, we account for the low-energy ($\rm{< 0.1~keV}$) photons 
that affect the abundances of key ions, such as C~\textsc{iv}, 
N~\textsc{v} and O~\textsc{vi}. 
We also calculate emergent spectra throughout the UV band, where 
the BAL features associated with the resonance transitions of these 
species are found.
 
\section{Method}
\label{method}
In this section, we will introduce the radiative transfer code we
use and review the geometry and kinematics of the outflow computed by
PK04. We also describe the form of the illuminating spectrum we adopt
and discuss how it relates to that used in PK04 and SP10.  

\subsection{\textsc{python} - a Monte Carlo radiative transfer and photoionization code}
We use the hybrid Monte Carlo / Sobolev code \textsc{python} for our calculations. 
This was originally described in  \cite{long_knigge} and updated in \cite{higg13}. Briefly, 
our simulation tracks photon-packets produced from a geometrically thin accretion disk, 
modelled as a series of blackbodies, and a central source of radiation that can be used to 
simulate an X-ray emitting corona. These packets are tracked through the wind, and their 
heating and cooling effect is computed. The ionization state is calculated using a model 
of the mean intensity in
 each cell of the simulation, calculated from the photon packets that pass through it. Once 
 a converged ionization state and temperature has been computed for the wind, synthetic 
 spectra can be produced for any required sightline.

The heating and cooling rates in a given cell include contributions from free-free, bound-free, 
bound-bound and Compton processes. Dielectronic recombination is accounted for as a bound-free
 cooling term. The ionization state in the cell is calculated under the assumption that the dominant 
 ionization process is photo-ionization from the ground state, which is balanced by radiative 
 recombinations from the ground state to all levels, plus dielectronic recombinations. Convergence 
 to the correct thermal and ionization equilibrium solutions is achieved by iteration, starting from
  an initial guess at the temperature and ionization state. In each iteration, or ``ionization cycle'',
   a representative population of photon-packets is tracked as it makes its way through the wind. 
   The energy deposited by these photons in each cell provides an estimate of the heating rate, 
   so the temperature of the cell can be adjusted to yield a more closely matching cooling rate. 
   The new temperature estimate is then used to obtain an updated estimate of the ionization 
   state of the cell. These updated cell properties then form the basis for the next ionization cycle.
    The calculations presented here were carried out using version 77 of \textsc{python} and include 
    H, He, C, N, O, Ne, Na, Mg, Al, Si, S, Ar, Ca and Fe.

\subsection{The PK04 disk wind geometry}
We take as our input the hydrodynamic wind model computed by PK04 at
time-step 955. This represents a time sufficiently far from the
simulation start that 
the wind geometry no longer depends on initial conditions. In
addition, at this point, the flow is only weakly time-dependent, so
this time-step is broadly representative of the flow as a whole. The
figures in PK04 also focused on this time-step, as did both SP10
and \cite{schurch_09}. The hydrodynamic calculations in PK04 
were carried out using \textsc{zeus} \citep{stone_norman}, modified to include the effect of 
the radiation field on the driving force and the temperature balance of the wind. 
The model is illustrated in Figure \ref{lin_den}, with the poloidal
component of velocity field superimposed upon 
the density structure. PK04 noted three main regions (delineated on Figure \ref{lin_den} by radial lines):
\begin{itemize}
\item{A: A hot, low-density, infalling region near the pole.}
\item{B: A hot transition zone in which the wind struggles to escape.}
\item{C: A warm, dense, equatorial flow that is shielded from the
  central source by a high-density `failed wind' region close to the
  origin} 

\end{itemize}

The \textsc{zeus} grid is polar, logarithmic in both the r- and $\rm{\theta}$-direction, 
with finer discretisation close to the
disk and the origin. To carry out the radiative transfer calculation, we create
 an identical grid and 
import the density and velocity structure of the \textsc{zeus} model. 
The density in each of our grid cells is constant and taken to be the value at 
the centre of each matching grid cell in the hydrodynamic model,
whilst the velocity field in each cell is interpolated across each cell
based upon the values at the vertices in the original model. This
discretization of the velocity field lends itself more naturally to
our use of the Sobolev approximation in the treatment of line scattering.

\begin{figure}
\includegraphics{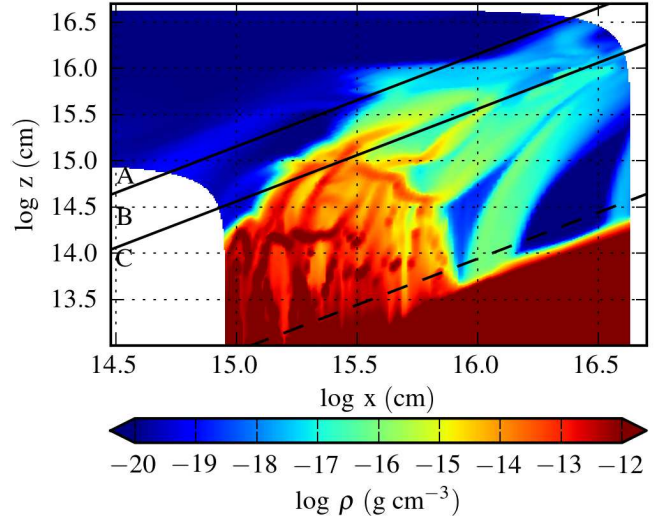}
\caption{The density of the PK04 model, plotted on logarithmic axes. The solid lines show the boundaries 
between the three zones described in the text, and the dashed line 
at $\theta=89.5^{\circ}$ is where we truncated the wind in our representation of the model.}
\label{density}
\end{figure}

Figure \ref{density} shows the same density information as Figure
\ref{lin_den}, but on logarithmic
axes in order to highlight the dense parts of the wind closer to the
disk plane. This shows a
wedge-like region at the base of the wind - the disk atmosphere. In our code, photons
emitted by the accretion disk start from the disk plane and therefore would have to make their way 
through this atmosphere before entering the rest of the wind. Such dense regions, in which photons 
scatter a very large number of times, are extremely computationally
expensive in a Monte-Carlo code. We therefore remove this part of the
flow from the computation, 
permitting photons to fly unimpeded from the disk into the lower part
of the wind. This is a reasonable approximation since, to first order,
the dense disk atmosphere would be expected to thermalise to the same
temperature as the disk beneath it. Hence the upper surface of that
region would be expected to emit photons with the same spectral
distribution as our disk.

The black line in Figure
\ref{density} shows the angle at which we make this cut - $\theta=89.5^\circ$.
The temperature of the lower surface of the flow where it is primarily
irradiated by photons from the disk is approximately equal to the disk
below it. This is the same as would be expected if the surface was in
direct contact with the upper surface of an extended disk
atmosphere. We are therefore confident that this simplification does
not significantly affect the results obtained for the bulk of the wind. 

\subsection{Illuminating spectrum}
Given the density and velocity structure of the wind, we next need to
specify the parameters of the radiation sources in our model, i.e. the
accretion disk and X-ray-emitting central `corona'. To permit
comparison between our results and those of PK04 and SP10, we
 match the parameters used in those studies as closely as possible.
\begin{figure}
\includegraphics{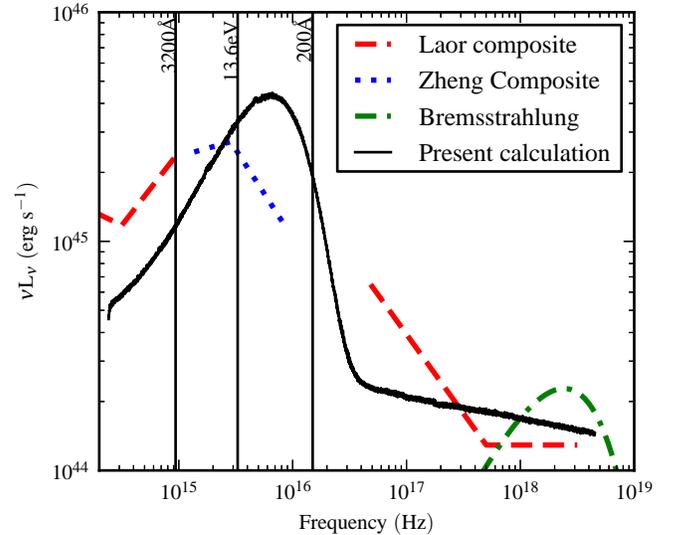}
\caption{The input spectrum using the simulation, together with the 10~keV bremsstrahlung spectrum
assumed in the original PK04 calculation. Also shown are the composite spectra from \citet{laor_97} and 
\citet{zheng_97} used to inform the luminosities used in PK04. The vertical lines illustrate the positions
of various band boundaries discussed in the text}
\label{input_spec}
\end{figure}

PK04 assumed the presence of a $\rm{10^{8}M_{\odot}}$ supermassive
BH at the centre of their accretion disk and adopted an
accretion rate on the BH of $\rm{1.8~M_{\odot}yr^{-1}}$. This gives a
disk luminosity of $\rm{L_D=6.1\times10^{45}~ergs~s^{-1}}$ assuming an
efficiency of $\simeq 6$\%.
In their hydrodynamic simulation, PK04 split the illuminating
luminosity into two components. First, they defined an `ionizing' luminosity
($\rm{L_{X}}$) that was used to compute the temperature and ionization
parameter throughout the wind. Second, they defined a `UV' luminosity
($\rm{L_{UV}}$) that was used to compute the actual line-driving
force. However, in regions of the wind that are shielded from the
X-ray source by optically thick material, the wind temperature was set
to that of the disk below the wind, thus implicitly assuming that
these parts of the wind would be heated and ionized by disk photons.

The balance between the two components was motivated by the composite
spectra presented in \citet{laor_97} and \citet{zheng_97}; the radio
quiet composite is shown in Figure \ref{input_spec}. 
The `ionizing luminosity', $\rm{L_X}$, defined by PK04 as the integrated luminosity of the
central source above 13.6~eV ($\rm{912~\AA}$), was taken to 
be 10\% of $\rm{L_D}$, or $\rm{6.1\times10^{44}~ergs~s^{-1}}$. The
ionization parameter at any point in the flow was then computed from
the local X-ray flux by assuming $1/r^2$ geometric dilution as well as
$e^{-\tau_{es}}$ attenuation, where $\tau_{es}$ is the optical depth
towards the origin due to electron scattering. This ionization was
then used to  calculate the line-driving force using the analytic formulae 
of \cite{stev_kall_90}. These formulae assume that the frequency
distribution of ionizing photons takes the form of a 10~keV
bremsstrahlung spectrum with no low-frequency cutoff. The calculations
of the wind temperature structure in PK04 used the same characteristic
temperature for the X-ray source in order to ensure
self-consistency. A bremsstrahlung spectrum of this form, normalised
to give the correct value of $\rm{L_X}$ is also plotted on Figure
\ref{input_spec}.

The UV luminosity in PK04 was defined as the luminosity in a band
running from 200 to 3000~$\rm{\AA}$ and was set to 90\% of $\rm{L_D}$. 
It should be noted that there is a degree of inconsistency here - a
significant part of the `UV' band is actually above the 13.6eV lower limit of
the `ionizing' band, and we see from Figure \ref{input_spec} that the
composite spectrum carries significant ionizing luminosity that would
contribute to $\rm{L_X}$, as well as $\rm{L_{UV}}$. However, increasing
$\rm{L_X}$ to take account of this would mean that the luminosity
around the peak of the bremsstrahlung spectrum would be much too
high. This illustrates the difficulty of approximating an observed SED
with a highly simplified model.

In our current calculation, we use a thin accretion disk with the same
BH mass and accretion rate adopted by PK04, even though such a model
may not produce the best match to observations \cite[e.g][]{davis_laor,done_12,sloan_netzer}. 
We also include a power-law X-ray source assumed to arise in a central spherical
`corona'. This X-ray source is defined by the same photon index, 
$\Gamma$, and luminosity in the 2-10~keV range, $L_{2-10}$, that was
used by SP10. In that work,
the X-ray parameters were chosen so as to give a reasonable match to
observations whilst maintaining the same luminosity in the 2-10~keV
range as the X-ray parameters in PK04. More specifically, we follow
SP10 in adopting $\rm{L_{2-10}=2.7\times10^{44}~ergs~s^{-1}}$ and
$\rm{\Gamma_{X}=2.1}$. 
Moreover, following PK04, we set the radius of
the X-ray source, $\rm{r_X}$, equal to the innermost radius of the
accretion disk (which we take to be the innermost stable circular orbit, 
$R_{ISCO} = 6 R_g$ for a Schwarzschild black hole, where $R_g$ is the gravitational radius). With
 these choices, our SED yields
$\rm{L_X\sim0.7L_D}$ and $\rm{L_{UV}\sim0.9L_D}$ in the notation of
PK04. Note the two components sum to more than unity because we count
ionizing photons from the disk in both $\rm{L_X}$ and
$\rm{L_{UV}}$. Figure \ref{input_spec} shows that our SED provides a
reasonable match to the composite spectrum derived from observations
and also produces roughly the same luminosity at high energies as the
bremsstrahlung spectrum. However, the ionizing luminosity in our model
is much higher than that in the PK04 simulation, since we allow both
disk and central source photons to contribute to the ionizing flux. We
return to this point in Section~\ref{IPdiff}, where we explicitly
check the impact of this difference on the calculated ionization state
of the wind. Our adopted parameters are summarised in Table
\ref{wind_param}. 

\begin{table}
\centering
\begin{tabular}{p{3cm}p{4cm}}
\hline Parameter 	&	 Value \\ 
\hline \hline 
$M_{BH}$ 	 &	 $10^8~\rm{M_{\odot}}$ \\ 
$\dot{M}_{acc}$ 	 &	 $1.8~M_{\odot}yr^{-1} \simeq 0.5~\dot{M}_{Edd}$\\ 
$\Gamma_X$ 	 &	 $2.1$ \\ 
$L_{2-10~eV} $ 	 &	 $2.7\times10^{44}~\rm{ergs~s^{-1}}$\\ 
$r_{disk}(min)=r_{X}$   &	 $6R_G=8.8\times10^{13}~{\rm cm}$ \\ 
$r_{disk}(max)$   &	 $2700R_G = 4\times10^{16}~{\rm cm}$ \\ 
\hline 
\end{tabular}
\caption{The key parameters adopted in the radiative transfer simulation.}
\label{wind_param}
\end{table}

\section{Results}
\label{results}
We are now ready to present the key properties of our converged wind
model. We will first illustrate the ionization state of the 
wind and then present the simulated emergent spectra. Finally, we will
also show the temperature structure and ionization parameter
throughout the outflow.

\subsection{Ionization state}

\begin{figure}
\includegraphics{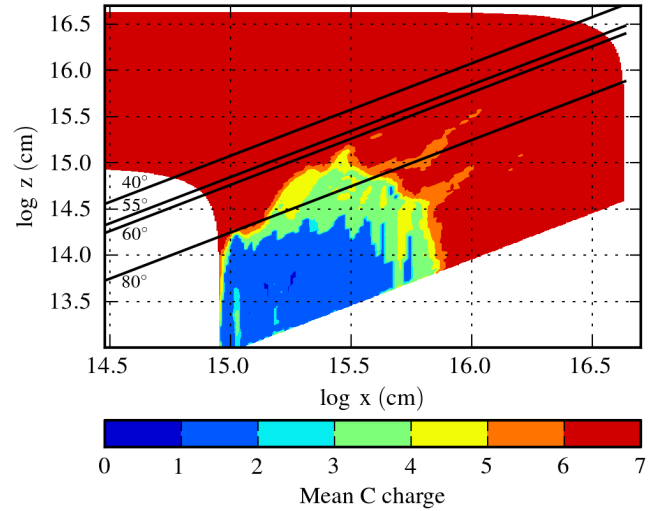}
\caption{The mean charge of carbon in the model.}
\label{ioniz}
\end{figure}

We begin by considering the ionization state of carbon, since strong,
blueshifted absorption associated with C~{\sc iv}~1550~\AA~ is a
defining feature in the UV spectra of essentially all
BALQSOs. However, as illustrated in Figure \ref{ioniz}, carbon is
actually fully ionized throughout much of the outflow. In the very dense 'failed wind' at the base of the wind, carbon is
in a much lower ionization state. This is partly due to the high
density in this region, but also to photoelectric absorption in the outer skin of the wind, 
which removes a large proportion of the soft X-ray photons responsible for ionizing carbon. In
PK04, the warm outflow behind the failed wind also remained in a low
ionization state. However, in our calculation, the carbon in this area
is fully ionized. This key difference is due to scattered and reprocessed
radiation and is discussed in more detail in section \ref{discussion}. Figure
\ref{ioniz} also illustrates the sight-lines for which we produce
synthetic spectra. 

\begin{figure}
\includegraphics{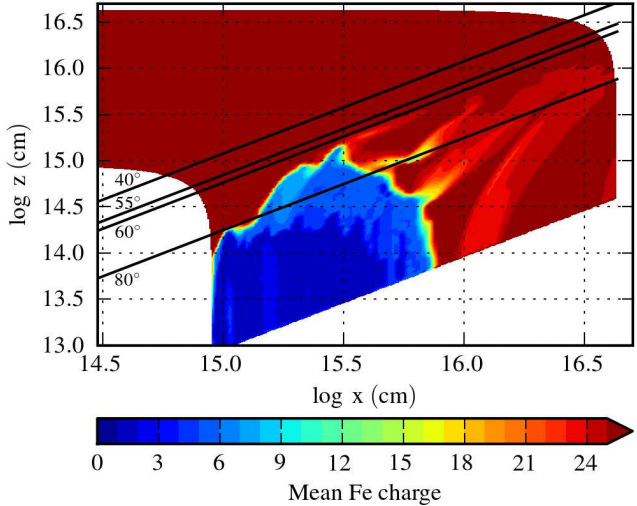}
\caption{The mean charge of iron in the model (c.f. \citealt{sim_proga_10}).}
\label{fe_ioniz}
\end{figure}

Figure \ref{fe_ioniz} shows the mean charge of iron in the model and
can be directly compared to Figure 3 in SP10. The computed ionization
states are very similar. This shows that the inclusion of an accretion disk in
our simulation as a source of ionizing photons has not altered the
conclusions of the earlier work with regards to species whose
ionization state is mainly set by the incident X-ray flux.   

\subsection{Synthetic spectra}

\begin{figure*}
\includegraphics{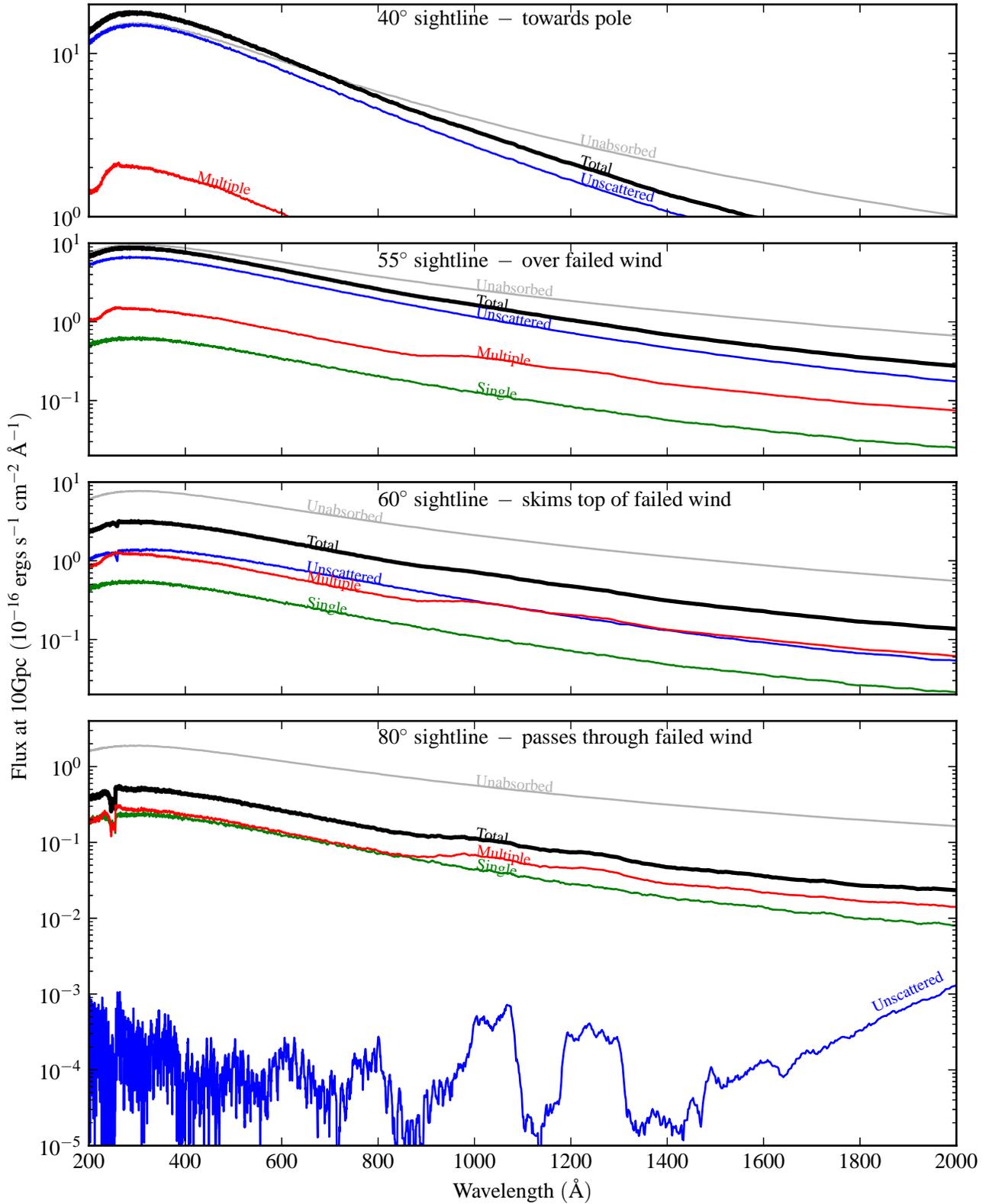}
\caption{The simulated spectrum for three sightlines. The grey line is the flux that would be
observed if there were no wind present. The heavy line shows the simulated spectrum for the converged
wind model, and the remaining three lines split that spectrum into components. The blue line shows the
part of the escaping flux associated with photons that have not scattered in the model, the green
line shows the part associated with photons that have scattered just
once, and the red line shows the part associated with multiply scattered photons.}
\label{spec}
\end{figure*}

Figure \ref{spec} shows synthetic spectra computed for the four sightlines illustrated in Figure \ref{ioniz}. For comparison, 
the spectrum that would be seen in the absence of any absorbing
material (labelled ``unabsorbed'') is also shown as a faint grey line 
for each sightline. The $\rm{80^\circ}$ sightline passes through the failed wind region where
carbon is not fully ionized. This is the 
most likely place to look for absorption from moderately ionized
species. However, the simulated spectrum for this sightline (the heavy
black line) is almost entirely
featureless except for a small absorption feature at $\rm{255~\AA}$ due to a resonance line of 
Fe~\textsc{xxiv}. The other three spectra in this plot show the
proportion of the net spectrum made up from unscattered, singly
scattered and multiply scattered photons. These contributions reveal why we do not see
any absorption features associated with the species present in the
failed wind: essentially all of the photons escaping along this
direction have scattered at least once and about 50\% of them have
scattered more than once. Thus the net spectrum for this sightline is
produced by photons that have scattered \emph{around} the failed wind, rather than
passing through it. Fe~\textsc{xxiv} is present in the equatorial, outflowing 
wind behind the failed wind region (see Figure \ref{fe_ioniz}) and
scattered photons pass through regions of the wind where this ion is
present. The direct (unscattered) component of the radiation field
escaping along this sightline is very heavily attenuated by a
combination of photoionization absorption and electron scattering.

The $\rm{60^{\circ}}$ sightline skims the top of the failed wind. Here, the component of the synthetic spectrum
associated with photons that did not undergo any scattering events is
only slightly less than 50\% of the total. As the viewing angle moves
even further up, to $\rm55^{\circ}$, most of the photons escape
unscattered. The Fe~\textsc{xxiv} feature is still present at
$\rm{60^{\circ}}$, but only in the direct component. The spectrum
predicted for the $\rm{55^{\circ}}$ sightline is completely
featureless. 

The final spectrum is computed for a sightline of $\rm{40^{\circ}}$,
and here we see a continuum enhancement at wavelengths below about
$\rm{700\AA}$. This is a consequence of the geometry of the wind 
model: photon packets with wavelength less than $\rm{700\AA}$ are predominantly 
produced by parts of the 
accretion disk inwards of the innermost radial extent of the wind ($\rm{60~R_G}$). 
The spectrum is therefore made
up of a direct component of photons passing unimpeded from the disk into the 
relatively low density
polar parts of the wind (direct component), and photons reflecting from the inner 
face of the failed wind (scattered 
component). By contrast,  a significant proportion of 
photon packets with wavelength longer than $\rm{700\AA}$ are produced from parts 
of the accretion disk under 
the dense failed wind, and this component will tend to be absorbed or scattered back 
into the disk. We therefore
only see a fraction of the direct component. Even when supplemented
by scattered radiation, the resulting flux is less than would be seen
in the wind were not present at all. The continuum enhancement at low
inclination angles was also predicted in the X-ray spectra computed in
SP10, where it was also attributed to reflection from the innermost
illuminated face of the `failed wind' region. 

\subsection{Temperature distribution}
\label{temp_dist}
\begin{figure*}
\includegraphics{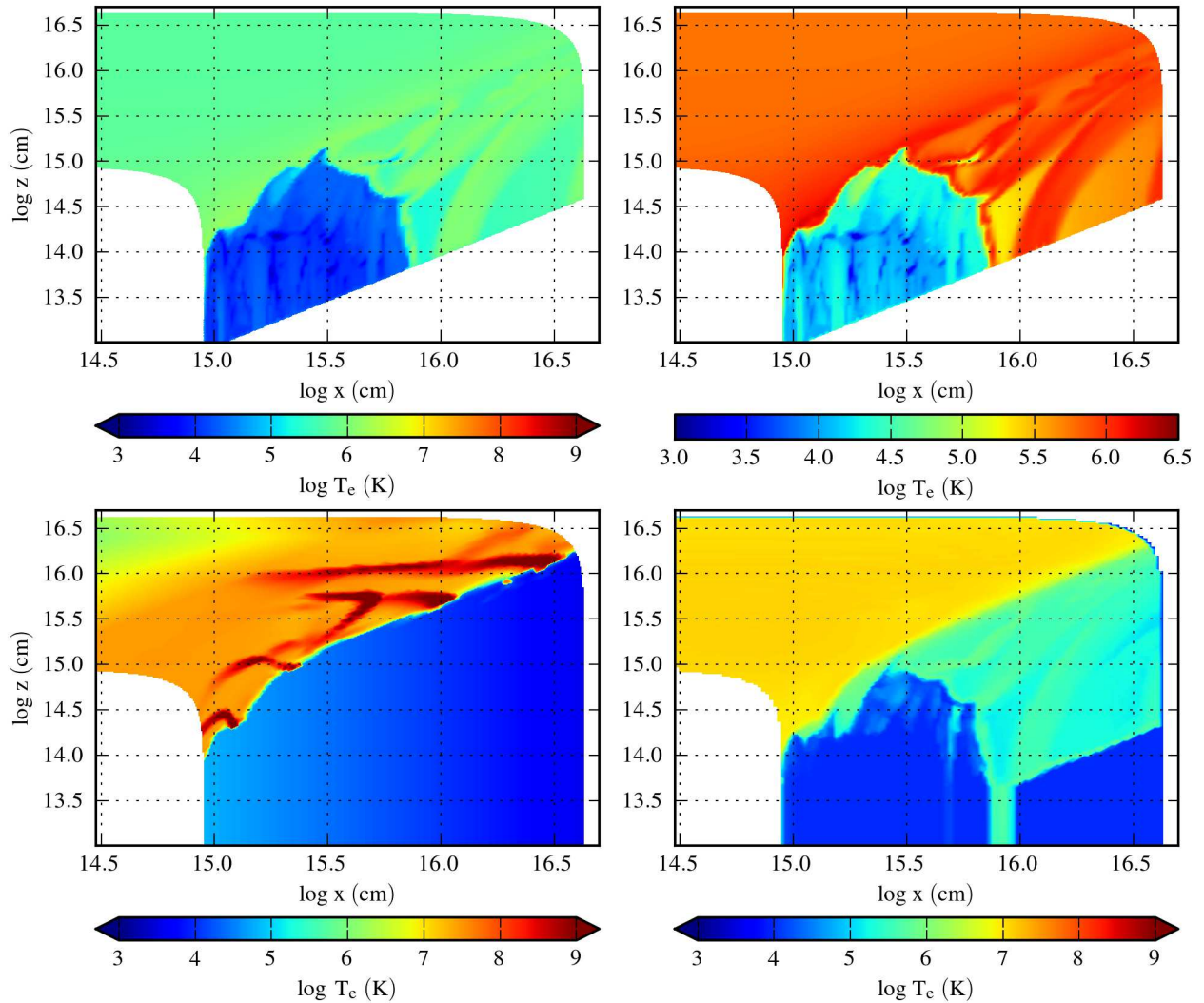}
\caption{The electron temperature ($\rm{T_e}$) of the model calculated here, compared to that from PK04 and SP10. The top
two panels show the current calculation, but on different colour scales. The upper right scale is optimised for the data, 
whilst the top left is on an identical scale to the lower two plots. The lower left panel is the temperature from the original 
PK04 calculation, whilst the lower right panel shows the temperature from SP10.}
\label{temp}
\end{figure*}

The top two panels of Figure \ref{temp} show the electron temperature
$T_e$ computed for the hydrodynamic model on two different scales. The
left hand matches the scale of the two lower plots, while the right hand
panel is optimised to show the full range of temperatures obtained in
the current calculations. The lower left panel shows the temperature
structure estimated by PK04, whilst the lower right hand panel shows
the temperature structure found by SP10. 
The clearest difference between the temperature computed in the
hydrodynamic simulation and that found here is in the low density
in-falling flow towards the polar direction. In this region, Compton
heating and cooling are the dominant mechanisms, and so the
temperature depends on the mean frequency of photons in the
region. Since we include the radiation field produced by the accretion
disk in our ionization and thermal equilibrium calculations of
ionization, the mean frequency we find is is far lower than that
assumed in PK04, where only the 10~keV Bremsstrahlung spectrum was
used to compute the ionization state and temperature structure. The
temperature structure found by SP10 is closer to ours, since they also
included an assumed disk spectrum in their calculation of Compton
cooling. In any case, the exact temperature in this polar region is
actually not all that important, since the density there is far too
low to imprint any spectral features. 

Another interesting difference is seen in the transition zone, where
the PK04 calculation contains very hot `streamers' of shock-heated
gas. We do not model this non-radiative heating mechanism, and so our
wind does show these structures. Again, this difference does not
affect the simulated spectra or our wider conclusions, since this
region is fully ionized in both cases.

Finally, a much more important difference is in the warm equatorial
flow. This region lies behind the dense `failed wind' region  
which is very cold all three calculations. In PK04, the ionization
parameter was used to calculate the temperature in the model. This, in
turn, was calculated using the X-ray luminosity of the central source,
attenuated by electron scattering opacity and distance. In the dense
equatorial flow, this estimate of the ionization parameter is very
low, and the treatment used in PK04 (described in detail in
\citealt{proga_stone_kallman}) yields unrealistically low
temperatures. In such cases, they assumed that the gas would be in
local thermodynamic equilibrium, and the temperature was set to be
equal to the temperature of the disk below it. However, our more
detailed 
radiative transfer treatment permits photons to scatter around the
failed wind, and these scattered photons turn out to be the dominant
source of ionization and heating in the equatorial flow region. We
therefore obtain temperatures of a few hundred thousand Kelvin for
this region, significantly hotter than in PK04, but similar to the
temperatures obtained by SP10, whose calculation also allowed for
scattering. We discuss the critical consequences of this finding in
Section~\ref{discussion}.

\subsection{Ionization parameter}

\begin{figure}[h]
\includegraphics{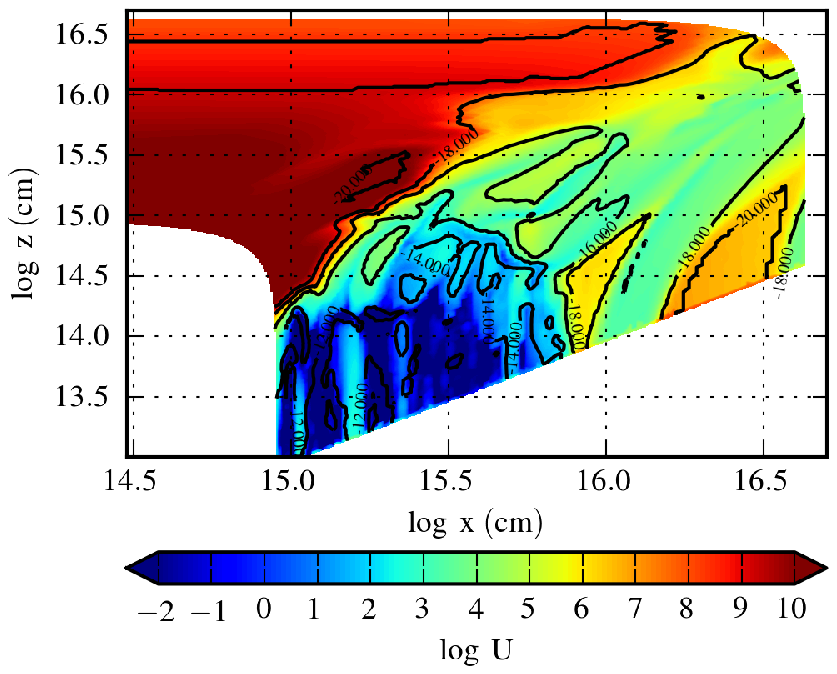}
\caption{The ionization parameter log(U) in each cell of the model. The contours show
the underlying density of the model.}
\label{IP_us}
\end{figure}

In Figure \ref{IP_us}, we show the ionization parameter $U$, computed
for each cell in the model based on the actual photon packets that
have passed through the cell. The standard definition of $U$ is
\begin{equation}
U=\frac{Q(H)}{4\pi r_0^2n(H)c},
\label{U_cloudy}
\end{equation}
where $\rm{Q(H)}$ is the number of photons emitted by the source per second, $\rm{r_0}$ is the distance from the 
illuminated cloud to the source, and $\rm{n(H)}$ is the number density of
hydrogen. Thus $U$ is simply the
ratio of the number density of ionizing photons, $\rm{n(\gamma_{H})}$,
to the number density of hydrogen atoms. We explicitly track this information in
our simulation, and so our version of $U$ for each cell is given by 
\begin{equation}
U=\frac{n(\gamma_{H})}{n(H)}.
\label{U_python}
\end{equation}
In the optically thin case, both definitions will give identical
values of $U$. However, our version self-consistently takes account of
any absorption between the source and the cell, along with additional
photons which may scatter into the cell from other lines of sight.

Figure~\ref{IP_us} shows immediately why we do not have any
C~\textsc{iv} in the wind. In order for a significant fraction of
Carbon ions to be found in this ionization stage, we require
$\rm{\log(U)\sim-2}$ \citep{higg13}. There are no parts of the wind
where this is true. There are also significant implications of this
ionization parameter distribution on the self-consistency of the
hydrodynamic model, since efficient line driving is only possible 
for significantly lower values of ionization parameter than we predict
for the outflowing parts of the simulation. We discuss this further
below.

\section{Discussion}
\label{discussion}

We are now in a position to consider some key aspects and consequences
of our results in more detail. First, we will attempt isolate the
cause of the discrepancy between the temperature structure and
ionization state found by PK04 and those calculated above. Second, we
will look at the way in which photons actually propagate through
the PK04 model in order to address what the structure of a viable
line-driven disk wind might be. Third and finally, we will discuss the
implications of our study for dynamical simulations of line-driven
winds.

\subsection{Why is the wind hotter and more ionized than predicted by PK04?}
\label{IPdiff}

There are several key differences between the calculation of the
temperature and ionization state of the wind in PK04 and that
described above: 

\begin{itemize}
\item{Input spectrum: PK04 used a 10~keV Bremsstrahlung spectrum
  to compute both the ionization parameter and the temperature of the
  wind; by contrast, we have used an observationally motivated AGN
  spectrum that includes ionizing radiation associated with both the
  accretion and a central X-ray source.}
\item{Wind radiation: in our model, photons produced by the wind
  itself are able to interact with (e.g. ionize) other parts of the
  outflow; in PK04, the wind was assumed to be transparent to its own
  cooling radiation.} 
\item{Scattering: in PK04, the local radiation field at each
  point in the wind was assumed to be dominated by {\em direct}
  radiation from the central engine, with electron scattering treated
  as a purely absorptive process; here we take full account of all
  opacity mechanisms and in particular we track the propagation of
  scattered photons through the flow.}
\end{itemize}

\begin{figure*}
\includegraphics{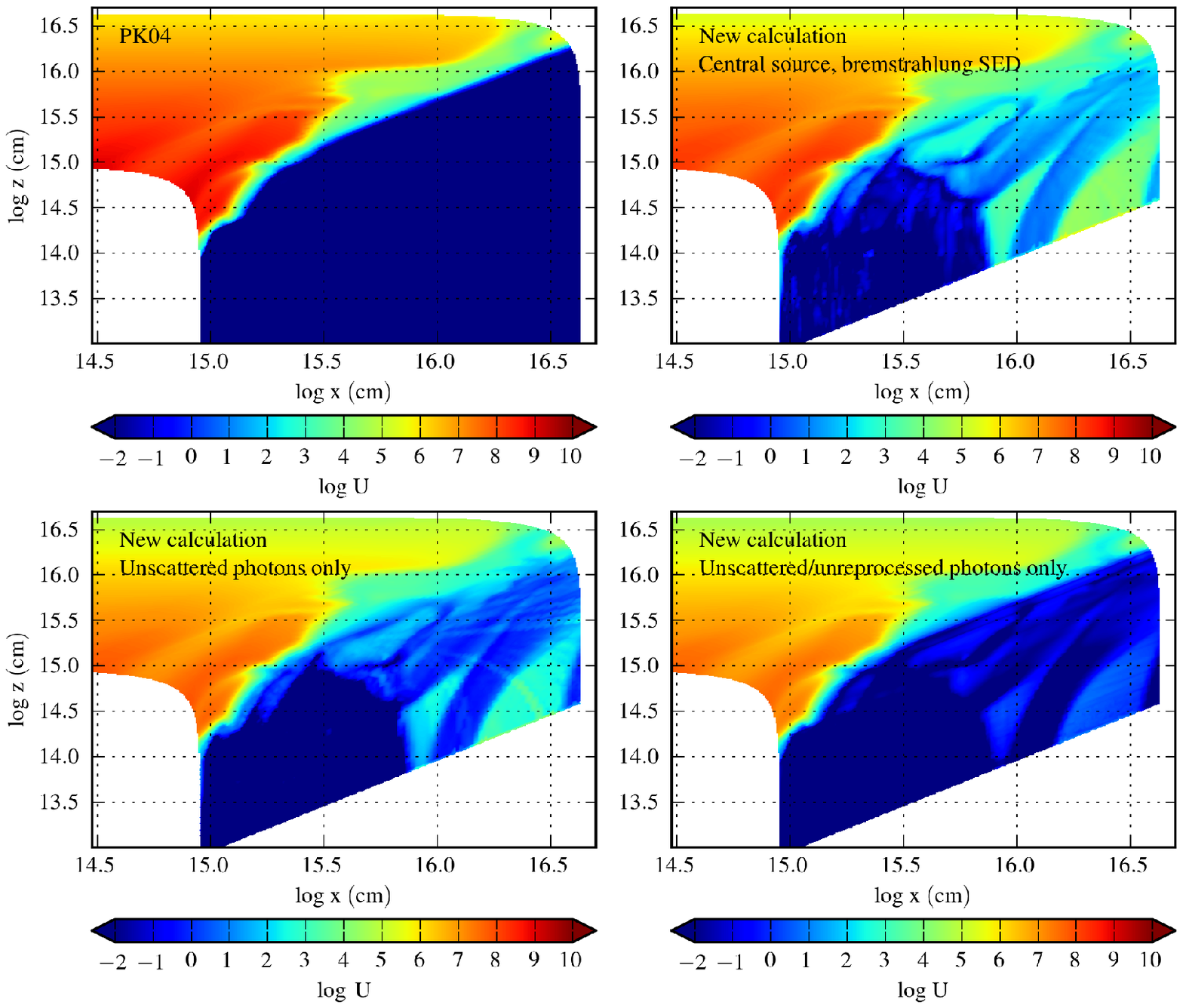}
\caption{The upper left panel shows the ionization parameter from PK04 
(scaled from their definition $\rm{\xi}$ to
ours via $\rm{\log(U)=\log(\xi)-1.75}$). The remaining three panels
all show results from \textsc{python} calculations that use an 
approximation to the 10~keV bremsstrahlung spectrum assumed in 
PK04 as the only source of photons. 
The
upper right panel shows $U$ calculated taking into account all photons in 
the cell, whilst the lower panel only counts photons arriving directly
from the X-ray source (i.e. unscattered photons). The lower right 
panel shows U from a calculation where the wind is permitted to
cool via radiative processes, but the resulting photons are not
actually produced and tracked in the simulation . This
is analogous to the approximation made in PK04. Here, again, 
only unscattered photons used to calculate $U$ (i.e. only unscattered
and unreprocessed photons).}
\label{IP_proga}
\end{figure*}

To assess the relative importance of these effects,
Figure~\ref{IP_proga} shows the distribution of the ionization
parameter throughout for two additional test cases. The ionisation
parameter is a useful parameter to compare here, since it largely
defines the ionization state of the wind and is also used to compute
the temperature and line-driving force in PK04. 

The upper right panel in Figure \ref{IP_proga} shows the ionization
parameter as presented in Figure 1 in PK04. However, we have converted
the definition of the ionization parameter adopted in PK04 ($\rm{\xi}$) to
that adopted here ($U$), using their approximate formula
$\rm{\log(U)=\log(\xi)-1.75}$. Comparing this to Figure \ref{IP_us},
we immediately see that our values of $U$ are consistently higher than
those found by PK04.  

To test how much of this difference is due simply to the
differences between the adopted illuminating spectra, we have rerun
our simulation, with the accretion disk removed as a radiation source
and with the X-ray spectrum modified to a power-law that approximates
the 10~keV Bremstrahlung spectrum assumed in PK04,
i.e. $F_{\nu}\propto\nu^{0}$. This X-ray spectrum is normalized so as
to give the same luminosity over the 2-10~keV range as the PK04 SED,
and the high-frequency cutoff of the Bremsstrahlung spectrum is
approximated by truncating the power law at 10~keV. This input
spectrum is used for all three of the new calculations shown in Figure
\ref{IP_proga}.  

The ionization parameter calculated for the wind in the absence of a 
disk radiation field is shown in the upper left panel of Figure \ref{IP_proga}. 
Based on this, we can conclude that the difference
between the ionizing SED adopted by us and in PK04 is the main reason
for the discrepancy between the estimates of $U$ in the polar wind
regions. This is  unsurprising since the wind is optically thin in
these directions. However, the critical region for the viability
line-driving and the production of broad UV absorption lines is the
warm, equatorial outflow. Here, $U$ is uniformly very small in the
PK04 data, but remains at $\log(U) \gtappeq 2$ outside the dense
failed wind region even in our simulation without disk photons. This
ionization parameter is too high to support line driving \citep{stev_kall_90}.

In the lower left panel of Figure \ref{IP_proga}, we show $U$ from the
same no-disk calculation, but taking into account only photons that
have not undergone any scattering events, i.e. only photons arriving
directly from the central source or emitted by the wind. We now see a 
moderate decrease in $U$ throughout the whole model except the bulk of
the failed wind. This is consistent with the failed wind efficiently
scattering photons into the polar regions in our standard calculation,
and with regions shielded by the failed wind being affected strongly
by radiation scattered around the shield. However, the ionization
parameter in the warm, equatorial outflow region is still higher than
in PK04 and still too high for efficient line driving. 

Finally, in PK04, the wind is able to cool via radiative processes,
but is assumed to be optically thin to its own radiation. We therefore
carry out another no-disk, unscattered-photons-only calculation, but
now also disable wind emission in the \textsc{python} simulation. More
specifically, the wind is still allowed to cool via radiative processes,
but no photons are produced that could interact with other parts of
the wind. The result is  shown in the lower right panel of
Figure~\ref{IP_proga}. Only in this last calculation, which most
closely mimics the approximations made in PK04, does $\log(U)$
drop to a level where line driving could become efficient. 

\subsection{The propagation of ionizing photons through the outflow: implications for line-driving}

We have seen that taking account of wind emission, multiple scattering
and including the accretion disk as a source of  
ionizing radiation has resulted in an outflow that is too highly
ionized for line driving to work efficiently as an acceleration  
mechanism. However, this does not mean that line driving could not
produce an outflow at all. If one were to compute a new step
in the hydrodynamical calculation based on our new ionization state,
the line-driving forces, and hence the outflow structure, would
certainly change (relative to the same time step in
PK04's original simulation). The key question is whether the end result
would be a total quenching of the wind or whether the flow might
adjust itself to a new configuration in which line driving can be
effective again. A self-consistent 3-dimensional
radiation-hydrodynamic calculation would be needed to answer this
question definitively, and we briefly comment on the prospects for
this in Section~\ref{line_driv} below. However, by considering the way
in which ionizing radiation actually propagates through the outflow,
we can already comment on the general properties of any viable
line-driven disk wind model.

\begin{figure}
\includegraphics{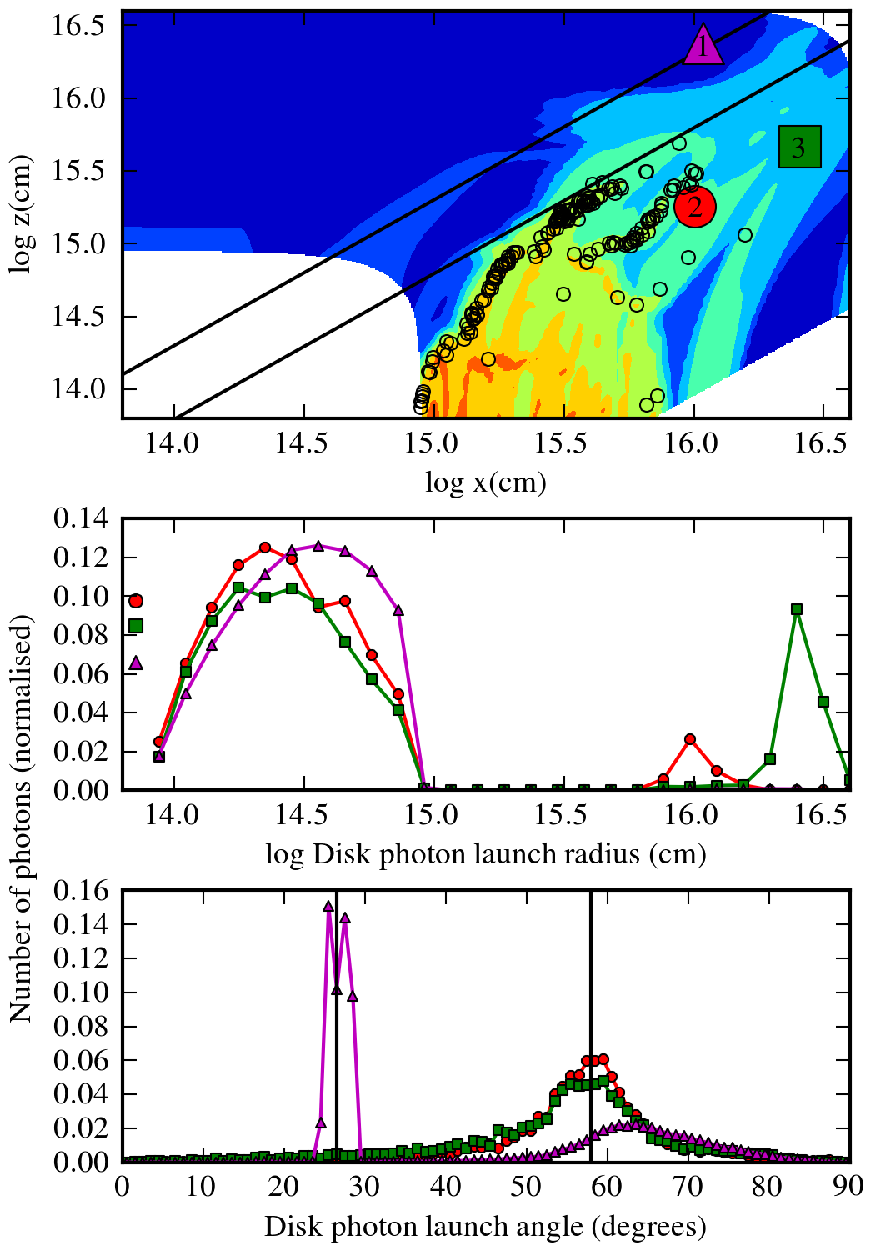}
\caption{Illustration of the source of ionizing photons contributing
to the radiation field in three cells. The top panel shows the
location of the three cells (filled symbols), along with the starting locations of wind generated 
photons that reach cell~2 (open circles). The underlying colours show
the density of the model - note that the colour scale here is different to
that used in Figure \ref{density}. The two straight lines in this plot
correspond to two specific sightlines: the top line marks $i \simeq
26.5^{\circ}$, the sightline passing from the origin through cell~1;
the bottom line marks $i \simeq 58^{\circ}$, a sightline passing
through the transition region between the failed wind and the polar
regions of the outflow. The middle panel shows the  launch
radius of the disk generated photons that reach the three cells
(plot symbols match the symbols used to mark each cell in the top
panel). The proportion of photons originating from the central X-ray
source is also shown via the isolated symbols near the left hand side
of the plot. The bottom panel shows the initial launch angle (relative to
the vertical axis) of the same photons. The vertical lines mark the
two angles also shown as sightlines in the top panel. In the lower two
plots, the ordinate is normalised to the total number of photons
reaching the cell in question.}
\label{phot_sources}
\end{figure}

In Figure~\ref{phot_sources}, we show some characteristics of the
ionizing photons that reach three particular cells in the wind. One
cell (1 - marked with the triangle in the upper panel) is in the highly
ionized, low density part of the model and serves as a 
comparison to the other two cells (2 - marked with a circle, and 3 - marked with a 
square), both of which are in the outflowing part of the model. In
PK04, cells 2 \& 3 are characterized by the relatively low ionization 
states required for efficient line driving. In our calculations, these
cells are highly ionized, so that line driving could not be effective
in them. Also shown in the upper panel are the starting locations of the ionizing
photons emitted by the wind that pass through cell 2. Broadly
speaking, these photons turn out to originate in the outside skin 
of the failed wind. This region sees and absorbs a significant
fraction of the ionizing radiation generated by the disk and the X-ray
source. Since the wind is assumed to be in radiative and thermal
equilibrium, all of this absorbed energy must be locally re-emitted. 

The middle panel in Figure~\ref{phot_sources} shows the distribution of
the initial launch radii of the ionizing disk photons that pass
through our test cells. The detached symbols near the left-hand-side
of the plot show the corresponding fraction of photons emitted by the
central X-ray source for comparison. We can see immediately that, for
all cells, the majority of ionizing photons arise in the inner parts
of the accretion disk. In PK04, the region above this part of the disk was outside the
computational domain, and so it is empty in our calculation as
well. Photons generated in this part of the disk are therefore able to
reach the wind without attenuation. In the hydrodynamic calculation, this was
done for reasons of computational efficiency, but it is also physically
reasonable, since the temperature and ionization state of the gas
above this part of the disk would likely be too high to permit efficient line
driving. Nonetheless, the additional failed wind region that would
likely result if this region was included in the simulation would
certainly modify the spectrum incident on the rest of the wind and
might result in a lower ionization state there. 
Extending the computational domain down to the innermost parts of
the disk would therefore be an important next step in understanding
line-driven winds. Useful steps towards this have already been taken
in radiation-(magneto)hydrodynamic simulations of accretion
disks \cite[e.g.][]{ohsuga_mineshige,jiang_13}.

The middle panel of Figure~\ref{phot_sources} also shows that the
launch radius distribution of ionizing photons is similar for all
three test cells, even though two of the cells (nos. 2 and 3) are
located in the shadow of the failed wind region, as seen from the
central engine. Thus the failed wind does not provide an effective
shield from ionizing radiation. As already shown in Figure~\ref{spec}
and discussed in Section~\ref{results}, this is because scattering
allows photons to reach parts of the flow that would otherwise be
shadowed. Interestingly, the radiation field in cells 2 and 3 also
includes some photons generated in the outer parts of
the disk; for example, about 10\% of the ionizing photons reaching 
cell~3 come directly from the disk beneath it. This is possible
because the outflow density and opacity has dropped significantly at
these larger radii.

Finally, the lower panel of Figure~\ref{phot_sources} shows the
distribution of launch angles (relative to the vertical axis) for all
ionizing photon packets produced by the disk and central X-ray source
that ultimately pass through the three test cells. The direct
sightline from the origin to cell~1 corresponds to an angle of
$\rm{26.5^{\circ}}$, and the launch angle of most photons that reach
this cell is, in fact, close to this value. However, a significant
fraction of the ionizing photons in this cell were originally emitted
in a more equatorial direction and only reach the cell after
scattering off the failed wind. These photons produce the hump near
$\rm{65^{\circ}}$ in the photon launch angle distribution for cell~1.
Cells 2 and 3 (which lie at a sightline from the origin of 
$\rm{80^{\circ}}$) are shielded from the direct radiation produced by
the central X-ray source and inner disk by the failed wind. The
ionizing photons reaching these cells instead exhibit a wide
distribution of initial launching angles, 
peaking around $\rm{58^{\circ}}$. This direction is marked by the
lower straight line in the top panel of
Figure~\ref{phot_sources}. Photons taking this initial path interact
with the hot transition region of the flow, i.e. the upper skin of the
failed wind. When they enter this region, the optical depth along
their trajectory is much higher than that perpendicular to it, so they
are preferentially scattered both up into the polar regions and down
into the parts of the flow behind the failed wind. Photons with
initially more equatorial directions will initially strike the failed
wind region. A significant fraction of these photons will be reflected
back across the inner disk and enter the transition region
on the other side of the grid. Some of these photons will then once
again be scattered in the outflow region behind the failed wind.  

This analysis suggests that it is the transition region (i.e. region B
in Figures \ref{lin_den} and \ref{density}) that is mainly
responsible for redirecting ionizing photons into the outflow regions
behind the failed wind. This applies not only to photons emitted by
the central X-ray source and inner disk, but also to photons produced 
by the ``front'' of the wind itself. Thus, even though \textit{the failed wind
is doing its job in preventing direct illumination} from reaching the
outflow behind it, ionizing photons can get around it via scattering
in the transition region. 

How might we expect the outflow to respond? Given the over-ionization
of the outflow behind the failed wind, it seems 
likely that the outflow in this region would also fail. This would
result in a larger failed wind region - perhaps reducing 
the ionization state of the material behind this extended block. However, 
since the failed wind is already highly optically thick in the radial 
direction, and since scattered and reprocessed radiation is already
dominant in the shielded region, this effect is likely to be quite
small. 

Given the importance of the transition region in redirecting ionizing
photons into otherwise shielded parts of the flow, it seems likely
that it is this region that would have to be modified to allow the
formation of a line-driven outflow. In the PK04 model, the geometry of
this transition region allows photons that would otherwise pass over
the outflow to be efficiently scattered down into it. If this transition 
region instead had a more curved upper surface profile, photons missing 
the failed wind could not be scattered so easily back down into the
outflow behind this region. Is it plausible that the flow might adjust
itself to form such a geometry? Possibly. In the PK04 model, line
driving already only just works in the transition region, since the
ionization state there is fairly high. Given the even higher
ionization state we predict for this region, line driving would likely
fail completely, causing the material to drop back to the disk,
perhaps forming the geometry required to produce an effective
shield. However, fully self-consistent calculations would be required
to test this idea.

\subsection{Implications for hydrodynamic simulations of line-driven winds}
\label{line_driv}

The analysis we have carried out on the PK04 geometry has clearly
shown that reprocessing and scattering effects have a significant --
and actually dominant -- effect on the ionization state of the
wind. This, in turn, must strongly affect the efficiency of the line
driving mechanism itself. The clear conclusion from our results is
therefore that these radiative effects must somehow be incorporated
into hydrodynamic simulations. 

The most obvious way to achieve this would be to carry out a full
radiative transfer and ionization calculations after each time step of
the hydrodynamic simulation. However, the main radiative 
transfer calculation presented here takes about 10 hours of processing time 
on 192 processors of the Southampton University Iridis
supercomputer. Thus, even if such a scheme could be
implemented, it is not at all clear that it is computationally
feasible at this time. Nonetheless, it is important to develop at
least an approximate method along these lines in order to obtain
robust results from line-driven wind simulations.

It should be noted that it is not only full hydrodynamic simulations
of line-driven winds that have treated radiative transfer in a
simplified manner. Calculations such as those presented in 
 \cite{ris_elv} and \cite{nomura_13} also neglect scattered and
reprocessed radiation and give rise to wind solutions similar to that
presented in PK04. It is therefore likely that similar effects would
be seen in these models.

\section{Conclusions}

We have presented results from comprehensive radiative transfer and
photoionization calculations for the line-driven AGN disk wind
predicted by the hydrodynamic simulations carried out in PK04. These 
simulations were computationally expensive and therefore 
treated the radiative transfer and ionization in a simplified
manner. Here, we have focused on one snapshot from the hydrodynamic
calculation and carried out a much more detailed Monte-Carlo
simulation of the interaction between the AGN radiation field and the
outflow.

Our main result is that the ionization state of the outflow is much
higher than estimated by PK04, to the extent that line-driving would
become inefficient. The over-ionized flow also no longer produces the
broad UV absorption lines that are the key observational tracers of
disk winds in AGN/QSOs. The main reason for this change in the
predicted ionization state of the flow is that self-shielding becomes
much less effective when radiative transfer effects are
fully accounted for. More specifically, the failed wind region that
protects the outflow from over-ionization in the simulations of PK04
can actually be ``circumnavigated'' by ionizing photons via scattering
and reprocessing. 

We conclude that hydrodynamic disk wind simulations need to take
account of scattering and reprocessing in order to robustly assess the
viability of line driving as an acceleration mechanism. This should
ideally take the form of a self-consistent treatment, in which the 
radiative transfer and hydrodynamics are calculated
simultaneously. However, the kind of `post-processing' approach we
have used here is at least useful in validating purely hydrodynamic
models. For outflows driven non-radiatively, it already provides a
means to self-consistently predict the observational characteristics
of the flow.

\section*{Acknowledgements} The work of NH, JHM and CK are supported by the Science
and Technology Facilities Council (STFC), via studentships and a
consolidated grant, respectively. DP acknowledges Support for Program number HST-AR-12150.01-A that was
provided by NASA through a grant from the Space Telescope Science
Institute, which is operated by the Association of Universities for
Research in Astronomy, Incorporated, under NASA contract NAS5-26555.
This work was also supported by NASA under Astrophysics Theory Program
grants NNX11AI96G.

\bibliographystyle{mn2e}

 \label{lastpage}

\end{document}